# Multimedia Application for Solving a Sudoku Game

**Alasu Paul Sabrin**
"Tibiscus" University of Timisoara, Romania
ghedeon@gmail.com

ABSTRACT: This article explains the way in which, with the help of Action Script 3 in combination with Flash, a method of solving Sudoku game was implemented, through searching for the certain numbers and after that trying to guess for the squares where there are two possible numbers.

**Introduction**

In the late 19th century, number puzzle first appeared in a newspaper, when La France puzzle setters began to try removing numbers from magic squares. In 1892 Le Siècle, a Paris-based daily, published a partially completed 9x9 magic square with 3x3 sub square. Still this puzzle was not the present Sudoku because it contained double-digit numbers and required arithmetic rather than logic to solve, but it shared key characteristics: each row, column and sub-square added up to the same number.

Le Siècle's rival of La France, refined the puzzle almost like the modern Sudoku. Every 9x9 magic square puzzle so that each row and column contained only the numbers 1-9. Although they are unmarked, each 3x3 sub-square dose not indeed comprise the numbers 1-9. However, the puzzle can't be considered the first Sudoku because, under the modern rules, it has two solutions. The modern puzzle setter ensured a unique solution with two diagonal from 1 to 9 numbers.

These weekly puzzles were a feature of newspaper titles including L'Echo de Paris for about a decade but disappeared about the time of the First World War.

According to Will Shortz, "the modern Sudoku was most likely designed anonymously by Howard Garns, a 74-year-old retired architect and freelance puzzle constructor from Indiana, and first published in 1979 by Dell Magazines as Number Place (the earliest known examples of modern Sudoku). Garns's name was always present on the list of contributors in issues of Dell Pencil Puzzles and Word Games that included Number Place, and was always absent from issues that did not. He died in 1989 before getting a chance to see his creation as a worldwide phenomenon. It is unclear if Garns was familiar with any of the French newspapers listed above."

Nikoli was the first man who introduced the Sudoku game in Japan in a paper in April 1984 as "Suuji wa dokushin ni kagiru", which can be translated as "the digits must be single" or "the digits are limited to one occurrence".

Later on the name of the game was changed by Maki Kaji in Sudoku. In 1986, Nikoli introduced to innovations: the puzzle become symmetrical (meaning the givens were distributed in rotationally symmetric cells) and the number of givens was restricted to no more than 32. It is now published in mainstream Japanese periodicals, such as the Asahi Shimbun.

## 1 Action script

ActionScript 3.0 is a powerful, object-oriented programming language that signifies an important step in the evolution of the capabilities of the Flash Player runtime. The motivation driving ActionScript 3.0 is to create a language ideally suited for rapidly building rich Internet applications, which have become an essential part of the web experience.

Earlier versions of ActionScript offered the power and flexibility required for creating truly engaging online experiences. ActionScript 3.0 now further advances the language, providing superb performance and ease of development to facilitate highly complex applications, large datasets, and object-oriented, reusable code bases. With ActionScript 3.0, developers can achieve excellent productivity and performance with content and applications that target Flash Player.

ActionScript 3.0 is based on ECMAScript, the international standardized programming language for scripting. ActionScript 3.0 is compliant with the ECMAScript Language Specification, Third Edition (ECMA-262). It also contains functionality based on ongoing work on ECMAScript Edition 4, occurring within the ECMA standards body.

ActionScript is executed by the ActionScript Virtual Machine (AVM) built into the Flash Player. AVM1, the virtual machine used to execute legacy ActionScript code, powers Flash Player today and makes possible a wide range of interactive media and rich Internet applications.

## 2  Flash

Since its introduction in 1996, Flash has grown in popularity to become widely regarded as a standard for high-end multimedia Web sites and presentations. Flash derived from other Macromedia applications, particularly FutureSplash and Director. Macromedia Director has a longer history, but is primarily used for multimedia development for CD-ROMs, movies and television. However, the files created by Director are too large to port effectively over the Web.

Flash offers many of the dynamic features that Director offers, yet Flash compresses file sizes, making it possible to offer media-rich content and fast download times.

Flash combines four elements that define its functionality: vector graphics, streaming capability, a timeline, and layers.

Flash uses vector graphics, rather than bitmapped graphics such as GIF, JPG or PNG. Vector graphics perform more efficiently on the Web because they are based on mathematical computations, rather than the pixel-by-pixel information used by bitmaps.

As such, Flash graphics are scalable without affecting file size. For example, suppose that the two circles in Figure 23-l are separate vector graphics being displayed in a browser. Both images would have the same file size. The only difference between the two circles is the radius, which can be adjusted by a mathematical calculation. Vector graphics use mathematics in this way to manipulate images.

## 3  Describing the application

The main purpose of the application is to solve any Sudoku game with a visual representation of the result and the solving method.

The visual part of the application is made in Flash and under the graphic component, i used Action script 3. In the first part, I create a class named Square.as, which contains the script for creating each square of the board and sets some properties for describing the status of the square and the actions performed to the square. The second class used in the application is Sudoku.as and this is the main class.

## 4 Square.as

This class contains a public function that extends a MovieClip Class. The constructor of the class:

```
public function Square()
{
    digit_txt.text = "";
    digit_txt.restrict = "1-9";

    digit_txt.addEventListener( Event.CHANGE, onChange );
    digit_txt.addEventListener( MouseEvent.CLICK, onClick );

    certain = true;
}
```

sets a restriction for square content and creates some listeners for different events like CHANGE and CLICK:

```
private function onClick( e:MouseEvent ):void
{
    digit_txt.text = "";
    _digit = 0;
}

private function onChange( e:Event ):void
{
    _digit = uint( digit_txt.text );
    if( digit == 0 )
    {
        digit_txt.text = "";
        certain = true;
    }
}
```

The first function onClick clears the square when a click is performed in the square and sets the square value, determined by the _digit variable, to 0. The second function onChange sets the square value when the user enters a number in the square and puts the certain variable on true. This certain variable is needed because the solving algorithm needs to remember the numbers before he starts guessing the solution.

The rest of the functions defined in the class are pure technical ones for making the class work properly.

## 5 Sudoku.as

This is the main class which contains the algorithm for solving the board.

The algorithm used is a relatively simple one. In the first part the algorithm tries to find all the one posible choices, exactly like in the real life when we try to solve a sudoku: we find the square where we know for sure what number to put. After finding all the squares with this property, the algorithm takes the first square where 2 numbers are possible, puts one of them in the square and tries to solve sudoku with it; if he reaches a conflict in a row, column or 3x3 square, it starts over with the second number in the square. This is why we need that certain variable to erase the numbers that are not 100% sure before we start over.

The public function creates the board, using the square constructor we created earlier, and the buttons' listeners.

```
public function Sudoku()
{
squares = [];
for( var ycoord:uint=0; ycoord<9; ycoord++ ){
      for( var xcoord:uint=0; xcoord<9; xcoord++ ){
            var s:Square = new Square();
            s.x = 30 + xcoord * 30;
            s.y = 30 + ycoord * 30;
            s.xcoord = xcoord;
            s.ycoord = ycoord;
            s.digit = uint( testSudoku[ycoord].substr( xcoord, 1
) );
            s.digit_txt.tabIndex = ( ycoord*9 ) + xcoord;
            addChild( s );
            squares.push( s );
            }
      }

      solve_btn.addEventListener( MouseEvent.CLICK, solve );
      clear_btn.addEventListener( MouseEvent.CLICK, clearGrid );
      restore_btn.addEventListener( MouseEvent.CLICK,
restoreGrid );
}
```

The restore_btn is a button which restores the board to a model predefined in an array like this:

```
private var testSudoku:Array = [ "-43-12---", "-----58--", "2--
39-41-", "496-3----", "5-24-17-3", "----8-564", "-68-74--5", "--
18-----", "---15-37-" ];
```

25

Another important function is solve which is the function called when the solve button is pushed.

```
private function solve( e:MouseEvent ):void
{
      if( timer != null ){ timer.stop(); }

      counter = 0;
      tries = 0;
      tolerance = 1;
      triesSinceLastResult = 0;
      certainMode = true;
      initialFreeSquares = freeSquares = countFreeSquares();

      timer = new Timer( 10 );
      timer.addEventListener( "timer", timerHandler );
      timer.start();
}
```

In the first place, some variables are defined. Tolerance is a variable, which gives us the number of numbers per square. countFreeSquares() is a function which counts the number of free squares on the board.

When the solve button is pressed he dispatches one event only, but we need to recall some functions in order to solve the board as many times as needed; for that we created a timer which dispatches an event every 10 milliseconds and so we are able to call some function at that time.

The timer will be stopped when the board is solved.

```
if( checkComplete() )
{
      timer.stop();
}
```

```
private function timerHandler(event:TimerEvent):void {
                 var nineSquares:Array = getNineSquares();

                 var mn:Array = getMissingNumbers( nineSquares );

                 for( var i:uint=0; i<mn.length; i++ ){
                          checkPossiblePositions( mn[i], nineSquares );
                          tries++;
                          triesSinceLastResult++;
                 }

                 if( freeSquares != 0 ){
                          if( triesSinceLastResult > ( 4*freeSquares ) ){
                                   clearUncertainSquares();
                                   freeSquares = countFreeSquares();
                                   tolerance = 2;
                          }
                 }

                 stats_txt.text = "Attempts: " + tries + ", succes rate: " + Math.round( (((81-freeSquares)-(81-initialFreeSquares))/tries)*100 ) + "%";

                 counter = (counter + 1) % 27;
        }
```

getNineSquares() function is a function that returns an array of nine square formation (row, column, square). For verifying if the number exists on a row, column or 3x3 square I created a function, which gets all the missing numbers, and after that I call the checkPossiblePosition function, which searches in a row, column or 3x3 square if the number repeats or not. If not, I set that number for the square. The counter variable is used for selecting which case of the formation is used (row, column or 3x3 square).

After the check, if the algorithm doesn't find any result in a number of tries, increase the tolerance, for searching 2 numbers on a square, and try again also clearing the uncertain squares.

```
private function clearUncertainSquares():void {
      for( var i:uint=0; i<squares.length; i++ ){
           if( squares[i].certain == false ){
                 squares[i].digit = 0;
                 squares[i].digitColor = 0x333333;
           }
      }
}
```

For a better visualization of the resolving algorithm, the numbers are colored in different colors like this:

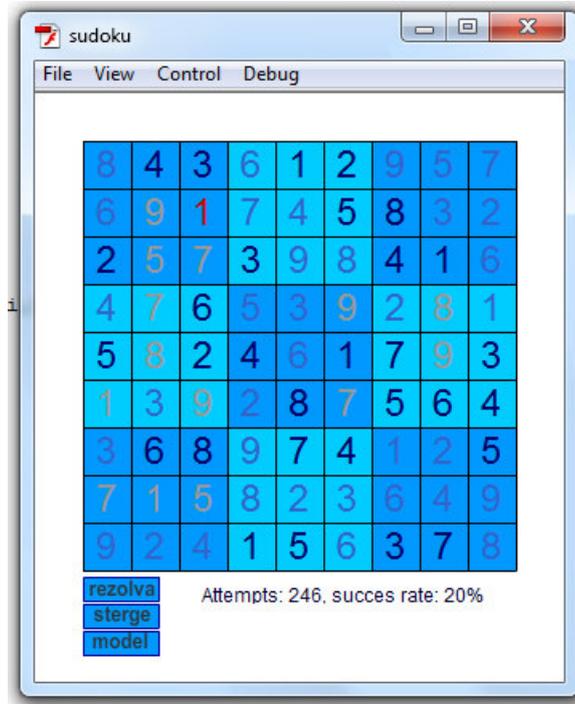

- The dark-blue ones are the model numbers (the numbers put in by the users)
- The light-blue ones are the certain numbers
- The grey ones are the uncertain numbers which are deleted if a conflict appears
- The red one is the number which is the computer guess (when the tolerance is 2)
- Attempts are the number of tries that the algorithm does for solving the board
- And success rate is the ratio of success in guessing the number

## 6  Rezults

In this part a series of tests were done for solving the board, with different grades of difficulty. The difficulty of the board is gained from the number of given fill squares:

- Easy: 34 – 35 filled squares
- Medium: 29 – 30 filled squares
- Hard: 26 – 27 filled squares
- Evil: 23 – 24 filled squares

| Nr Test | Difficulty | Squuares | Attempts |
|---|---|---|---|
| 1 | Easy | 34 | 213 |
| 2 | Easy | 34 | 78 |
| 3 | Easy | 34 | 115 |
| 4 | Easy | 35 | 127 |
| 5 | Easy | 35 | 81 |
| 6 | Medium | 29 | 371 |
| 7 | Medium | 29 | 166 |
| 8 | Medium | 30 | 450 |
| 9 | Medium | 30 | 230 |
| 10 | Medium | 30 | 167 |
| 11 | Hard | 26 | 2730 |
| 12 | Hard | 26 | 561 |
| 13 | Hard | 26 | 571 |
| 14 | Hard | 26 | 1123 |
| 15 | Hard | 27 | 1854 |
| 16 | Evil | 23 | 13674 |
| 17 | Evil | 23 | 28807 |
| 18 | Evil | 24 | 4732 |
| 19 | Evil | 23 | 45698 |
| 20 | Evil | 24 | 34895 |